# Approximation Algorithms for P2P Orienteering and Stochastic Vehicle Routing Problem

**Abstract**

We consider the P2P orienteering problem on general metrics and present a (2+ε) approximation algorithm. In the stochastic P2P orienteering problem we are given a metric and each node has a fixed reward and random size. The goal is to devise a strategy for visiting the nodes so as to maximize the expected value of the reward without violating the budget constraints. We present an approximation algorithm for the non-adaptive variant of the P2P Stochastic orienteering. As an implication of the approximation to the stochastic P2P orienteering problem, we define a stochastic vehicle routing problem with time-windows and present a constant factor approximation solution.

## 1   Introduction

The orienteering problem is a Prize Collecting-Travelling Salesman Problem (PC-TSP). In the P2P orienteering problem [3], given an edge weighted graph G= (V, E), two nodes u, v; the target is to find a u-v walk in G that collects the maximum reward subject to the constraint that the total distance travelled is less than a specified bound B. A node might be visited twice by the walk but is counted only once in the objective function. The orienteering problem is NP- Hard via reduction from the TSP. In the P2P Knapsack Orienteering problem (called P2P KnapOrient), there are two budgets i.e. a travel budget and a knapsack budget and the target is to find a u-v path with maximum reward such that the two budget constraints are met.

In the stochastic orienteering problem, we are given a metric, where each node has a deterministic reward and a random size. The goal is to decide which nodes to visit in order to maximize the expected value of the reward subject to the constraint that the cost of travelling plus the cost of processing the jobs is atmost B. One of the main motivations for the budgeted TSP problems comes from real life problems that arise in transportation, goods distribution etc. In the Vehicle Routing Problem with time windows, each node has an associated time window. The goal is to maximize the reward collected by visiting the nodes within their time windows. If deadlines of all the nodes are the same and the release times zero, the problem reduces to the orienteering problem. We consider the Stochastic Vehicle Routing Problem with Time Windows (SVRPTW) where each node has a random waiting/processing time along with a fixed reward. The VRPTW with stochastic waiting times encapsulates real life situations where the amount of time spent at a node (location) may vary.

### 1.1   Related Work

Blum et al. [2] gave the first constant factor approximation algorithm for the rooted Orienteering problem with an approximation guarantee of 4 via a (2+ε) approximation to the Minimum Excess path problem. In the same paper the problem is shown to be NP-Hard to approximate to within a factor of 1481/1480. This was improved to a 3 approximation for a stronger version of the problem called the P2P orienteering in [3]. The approximation for the minimum excess problem in [2] is achieved via the approximation for the k-stroll problem in [4]. The approximation ratio of 3 in [3] was improved to (2 + δ) in [1] via bi-criteria approximations for the minimum excess path and the k-stroll problems. The P2P algorithm of [3] requires O ($n^2$) applications of the approximation algorithm for the min-excess problem.

Gupta et al. [5] consider the stochastic orienteering problem and present a constant factor approximation algorithm for the best non-adaptive policy. They also demonstrate a small adaptivity gap –i.e. the existence of a non-adaptive policy whose reward is at least a Ω (1/log logB) fraction of the optimal reward and hence obtain an O (log logB) approximation for the adaptive problem.

The vehicle routing problem with time windows has been studied extensively in operations research and several heuristics have been studied to solve the problem optimally. In the approximation algorithms literature Chekuri and Kumar [6] gave constant factor approximation for the case when the number of different deadlines is constant. Bansal et al. [3] presented the first algorithm with approximation guarantees for the general case with arbitrary time-windows.

## 2   Notations and Preliminaries

The following notations are valid throughout the paper. Let G = (V , E) be a weighted graph , with a start node r, a prize (or reward) function $\prod : V \rightarrow \mathbb{Z}^+$, deadlines $D : V \rightarrow \mathbb{Z}^+$, release dates $R:V \rightarrow \mathbb{Z}^+$, and a length function $l:E \rightarrow \mathbb{Z}^+$. For any two nodes u,v ; l(u,v) denotes the shortest path between u and v ; $t_p(u,v)$ denotes the distance from u to v along P. The excess along a path P is defined as $\varepsilon_p(u,v) = t_p(u,v) - l(u,v)$. In the minimum excess problem, given a graph G, two nodes u,v and prize threshold k, the target is to find a u-v path that collects at least k prize and has the minimum excess. We assume that the deadline of every node is larger than the release date of the node and the shortest path distance from root to the node. A path may visit a node multiple times but the waiting time and the rewards are considered only once. The goal of the stochastic vehicle routing with time window (SVRPTW) problem is to construct a path, P, which maximizes the total reward collected by visiting the nodes within their time windows [R(v) ,D(v)].

The P2P algorithm in [3] requires $O(n^2)$ applications of the algorithm for the min-excess problem. Chekuri *et al.* [1] give a $(1/(1-\delta'), 2)$ bi-criteria approximation for the minimum excess problem and then use it to show the existence of $(2+\delta)$ approximation for the undirected orienteering problem. We present the P2P algorithm that requires $O(n)$ applications of the bi-criteria approximation algorithm for the minimum excess problem from [1].

In the knapsack orienteering (KnapOrient) problem [5], we are given two budgets L (travel budget), W (Knapsack budget), and a start node r. Each node has an associated reward $r_v$ and a size $s_v$. A feasible solution is a path P of length at most L, such that the total size $s(P) = \sum_{v \in V} s(v)$ is at most W. The goal is to find a feasible solution of maximum reward $\sum_{v \in V} r(v)$. In the P2P KnapOrient problem we are given two vertices u, v and the goal is to find a feasible u-v path that maximizes the reward.

An instance of stochastic orienteering is defined in [5] on an underlying metric space (V, d) with ground set |V| = n and symmetric integer distances d: VxV $\rightarrow$ $\mathbb{Z}^+$ (satisfying the triangle inequality) that represents the travel times. Each vertex is associated with a unique stochastic job which we also call v. Each job v has a fixed reward size $r_v \in Z_{\geq 0}$ and a random processing time/size $s_v$, which is distributed according to a known but arbitrary probability distribution $\pi_v$: R+ $\rightarrow$[0, 1]. In the P2P stochastic orienteering instance we are also given two nodes u,v and a budget B on the total time available. The goal is to devise a strategy, which starts at u, must decide (possibly adaptively) which jobs to travel to and process so as to maximize the expected sum of rewards of jobs successfully completed before the total time reaches its threshold B ,ending at v.

## 3  P2P Orienteering

In this section we present approximation algorithms for the both deterministic as well as the stochastic variants of the P2P orienteering problem. The knapsack orienteering (KnapOrient) problem defined and solved in [5] by putting the knapsack constraint into the objective function by considering the Lagrangian relaxation of the knapsack constraint. We solve the P2P KnapOrient problem by essentially using the approach and analysis used in [5]. The key difference being that we use the P2P algorithm presented in 3.1 as a subroutine. The P2P KnapOrient algorithm is used as a subroutine in the algorithm for the P2P stochastic orienteering problem.

### 3.1  Point-To-Point (P2P) Orienteering

P2P Algorithm

1) For every nodes x, we consider two u-v paths a and b.
a) Proceeds directly from u to x, then visits some vertices while travelling from x to v,
b) Visits some nodes while travelling from u to x, and then proceeds directly from x to v. The allowed excess of the indirect sections is $\varepsilon_x = D - l(u, x) - l(x, v)$. Applying the $(1/(1-\delta'), 2)$ bi-criteria approximation for the minimum excess problem from [1] on the indirect sections of the u-v path, we obtain paths with excess at most $\varepsilon_x$. If the excess along both indirect sections is less than $\varepsilon_x$, choose the path with greater reward. The path computed visits at least $(1 - \delta')$ fraction of the vertices visited by the best path on the indirect sections.

2) Now pick the x that maximizes the reward collected on the computed path.

The above algorithm requires $O(n)$ applications of the minimum excess bi-criteria approximation. Using analogy from the proof in [3], we show that P2P algorithm is a $(2 + \delta)$ approximation to the P2P orienteering problem. It is inbuilt in the algorithm proposed that the budget D is not exceeded, we ensure that the path returned visits sufficient vertices. Consider the optimum path O from u to v. Break the path at a point x, s.t. the two resulting sections have at least half the reward. Let the section with lesser excess be denoted by X and $\varepsilon_o(X)$ be the corresponding excess. Now considering path O' that follows O for the section X and picks the shortest path for the other half. Because we chose X such that $\varepsilon_o(X)$ is the smallest, we save atleast $\varepsilon_o(X)$ of excess in the other half; that is $t_o(u,v) - t_{o'}(u,v) \geq \varepsilon_o(X)$. By definition $\varepsilon_x = t_o(u,v) - t_{o'}(u,v) + \varepsilon_o(X)$; substituting the above inequality, $\varepsilon_x$ is atleast 2 $\varepsilon_o(X)$. Since the $(1/(1-\delta'), 2)$ bi-criteria approximation visits at least $(1 - \delta')$ fraction of the vertices visited by the best path in X, we are guaranteed $2/(1-\delta')$ [= $(2+\delta)$] fraction of the optimum reward.

### 3.2  P2P Knapsack Orienteering

**Theorem 3.1[5]:** There is a polynomial time O (1) approximation AlgKO for the KnapOrient problem. The AlgKO of [5] pushes the knapsack constraint into the objective function by considering a Lagrangian relaxation of the knapsack constraint. The rewards are altered while optimizing over the set of valid orienteering tours. An exhaustive search is used for obtaining the suitable lagrangian multiplier; a solution with large reward that meets both the knapsack and orienteering constraints is obtained via the (2+δ) approximation for the orienteering problem from [1]. The P2P knapsack orienteering (P2P KnapOrient) considered essentially uses the same approach and analysis as AlgKO. The difference being that the algorithm P2P-AlgKO uses the P2P algorithm presented in section 3.1.

### 3.3 P2P Stochastic Orienteering

We present an algorithm for the stochastic P2P orienteering that is essentially the same as the one presented in [5] with a slight modification. Let Opt be the optimal solution to the original P2P KnapOrient instance. The natural idea of replacing the random jobs by deterministic ones, with size equal to the expected size $E[S_v]$ , to find a near optimal orienteering solution P to the deterministic instance returns only an O[log B] approximation. The deterministic instances encode the mean but not the variance. The algorithms proposed in [5] is has the following structure theorem at its centre via truncated means and valid KnapOrient instances.

**Theorem 3.2[5]**: Given an instance $I_{so}$ for which an optimal non-adaptive strategy has an expected reward of Opt; either there is a single vertex tour with expected reward Ω (Opt), or there exists $W = B/2^i$ for some $i \in Z_{\geq 0}$ (or W = 0) for which the valid KnapOrient instance $I_{KO}$ (W) has reward Ω (Opt).

| Algorithm P2PAlgSO for P2P StocOrient on input Iso = $(V, d, \{(\pi_u, r_u) : \forall u \in V \}, B, \rho, \lambda)$ |
| --- |
| 1. for all $v \in V$ do |
| 2.    let $R_v = r_v . Pr_{S_v \sim \pi_v} [S_v \leq (B - d (\rho, v) - d (v, \lambda))]$ be the expected reward of the single vertex tour to v. |
| 3. w.p. ½, just visit the vertex v with the highest $R_v$, and exit. |
| 4. for $W = B, B/2, B/4...B/2^{\lceil log B \rceil}$, 0 do |
| 5.    let i = log (B/W) if W≠ 0, otherwise let i = $\lceil$log B$\rceil$ + 1. |
| 6.    let $P_i$ be the path returned by P2P-AlgKO on the valid KnapOrient instance $I_{ko}$ (W) |
| 7.    let $R_i$ be the reward of the P2P KnapOrient solution $P_i$. |
| 8. let $P_{i*}$ be the solution among $\{P_i\}_{i \in \lceil log B \rceil + 1}$ with maximum reward $R_i$. |
| 9. sample each vertex in $P_i$ independently w.p. ¼ & visit these sampled vertices in order given by $P_{i*}$. |

The above algorithm gives a randomized non-adaptive policy; it chooses a random path from ρ to λ to follow and just visits all the jobs until the budget allocated is exhausted.

## 4 A Stochastic Vehicle Routing Problem

In this section, we consider the Stochastic Vehicle Routing Problem with Time Windows (SVRPTW) and claim that the bi-criteria approximation algorithm proposed in [3] for the deterministic version of the same problem can be used to obtain a constant fraction of the optimal reward. In the SVRPTW, every vertex $v \in V$ is associated with a stochastic waiting (processing) time which is distributed according to a known but arbitrary probability distribution. Therefore when a path arrives at a node v, it must wait for a duration that equals the instantiation of the stochastic waiting time. In order to claim the reward associated with a node, it is essential that the waiting time ends within the corresponding time window.

**Theorem 4.1 [3]**: For any ε (> 0), there exists a polynomial time algorithm that obtains a $\frac{1}{24(s+2)}$ = Ω (log $^{-1}$ 1/ε) fraction of the reward obtained by the optimal path while exceeding the deadlines by a small fraction.

Bansal et al. use the 3 approximation algorithm to the P2P orienteering as a subroutine in the approximation to the Time-Window problem. The small and the large margin cases are considered separately and combined to obtain an approximation for the general case. In the small margin case, most vertices are visited by the optimal path very close to the deadlines. In the large margin case, the optimal path visits most vertices well before their deadlines. We mention the definition of only the general case of node splitting.

Let $f = 1/\sqrt{(1+\varepsilon)}$. Let s be defined as the smallest integer for which $f^{(1.5)^{\wedge}s} \leq$ ¼. Then s = O (log 1/ε). Divide the nodes into (s + 2) groups as follows. For $1 \leq i \leq s$ , $V_i = \{v : t_o(v) \in (f^{(1.5)^{\wedge}i} D(v), f^{(1.5)^{\wedge}(i-1)} D(v)]\}$. $V_0 = \{v: t_o (v) \in (fD(v), D(v)]\}$ . Groups (s+1) is defined as, $V_{(s+1)} = \{v: t_o (v) \in (0, D (v)/4]\}$.

The algorithm for the small margin case guarantees a 1/9 fraction of the optimal reward while the algorithm for the large margin case guarantees a 1/24 fraction of the optimal reward. In both the approximations, a factor of 3 is lost by the P2P orienteering subroutine.

An algorithm, essentially the same as the one proposed by Bansal et al. can be used to obtain a constant fraction of the reward for a SVRPTW instance. The difference being that instead of using the P2P subroutine, the stochastic P2P orienteering algorithm (AlgP2PSO) presented in 3.3 is used. The AlgP2PSO returns an O (1) fraction of the optimal reward. Therefore we obtain an O (1) fraction of the optimal reward for the SVRPTW instance while exceeding the deadlines by a small fraction. The above approach works because the approximation is done over the points visited and hence the corresponding reward and not on the time spent which varies according to the different instantiations of the waiting times at the nodes.

## 5  Conclusions

We present a (2 + δ) approximation algorithm for the P2P orienteering problem and a constant factor approximation for its stochastic version. We use this result to present constant factor bi-criteria approximation for a stochastic vehicle routing problem. The natural open question that arises is to improve these approximation guarantees. Also an important problem would be to reduce the running times of the algorithms without worsening the approximation guarantees.